\documentclass[aps,twocolumn,preprintnumbers,nofootinbib,superscriptaddress]{revtex4-1}
\usepackage{amsfonts}
\usepackage{amsmath}
\usepackage{amssymb}
\usepackage{graphicx}%
\usepackage{times}
\usepackage{amsmath}
\usepackage{amssymb}
\usepackage{graphicx}
\usepackage{subfigure,accents}
\usepackage{booktabs}
\usepackage{rotating}
\usepackage{longtable}
\usepackage{bm}
\usepackage[pagebackref=false,colorlinks=,linkcolor=blue,citecolor=blue]{hyperref}
\usepackage{epstopdf}
 \pdfoutput=1
 \usepackage[utf8]{inputenc}
\usepackage{cases}
\usepackage{amsmath,float}
\usepackage{amssymb}
\usepackage{amsfonts}
\usepackage{amssymb}
\usepackage{dcolumn}
\usepackage{bm}
\usepackage{bbm}
\usepackage{graphicx}
\usepackage{xcolor}
\usepackage{array}
\usepackage{subfigure}
\usepackage{hyperref}
\usepackage{wasysym}

\newcommand{\be}{\begin{equation}}
\newcommand{\ee}{\end{equation}}
\newcommand{\ba}{\begin{eqnarray}}
\newcommand{\ea}{\end{eqnarray}}

\newcommand{\gsim}{\mathrel{\hbox{\rlap{\lower.55ex \hbox {$\sim$}}
                   \kern-.3em \raise.4ex \hbox{$>$}}}}
\newcommand{\lsim}{\mathrel{\hbox{\rlap{\lower.55ex \hbox {$\sim$}}
                   \kern-.3em \raise.4ex \hbox{$<$}}}}

\hypersetup{colorlinks=true,
            breaklinks=true,
            pdfstartview=Fit,
            linkcolor=blue,
            citecolor=blue,
            urlcolor=blue}
            
 \usepackage{graphicx}

\usepackage{amsmath}

\usepackage{braket}

\newcommand{\bw}{\begin{widetext}}
\newcommand{\ew}{\end{widetext}}

\def\ber{\begin{eqnarray}}
\def\eer{\end{eqnarray}}
\def\beq{\begin{equation}}
\def\eeq{\end{equation}}

\begin{document}

\title{Avoidance of singularity during the gravitational collapse with string
T-duality effects}
\author{Kimet Jusufi}
\email{kimet.jusufi@unite.edu.mk}
\affiliation{Physics Department, State University of Tetovo, Ilinden Street nn, 1200, Tetovo, North Macedonia}

\begin{abstract}
    In this paper, we explore the gravitational collapse of matter (dust) under the effect of zero-point length $l_0$. During the gravitational collapse, we have neglected the backreaction effect of the pre-Hawking radiation (in the sense that it is small effect and cannot prevent the formation of apparent horizon), then we recast the internal metric of a collapsing star as a closed FRW universe for any spherically symmetric case and, finally, we obtain the minimal value for the scale factor meaning that the particles never hit the singularity.  We argue that the object emerging at the end of the gravitational collapse can be interpreted as Planck stars (black hole core) hidden inside the event horizon of the black hole with radius proportional to $(GMl_0^{2}/c^2)^{1/3}$. Quite interestingly, we found the same result for radius of the Planck star using a free falling observer point of view. In addition, we pointed out a correspondence between the modified Friedmann’s equations in loop quantum gravity and the modified Friedmann’s equation in string T-duality.  In the end, we discuss two possibilities regarding the final stage of the black hole. The first possibility is that we end up with a Planck-size black hole remnants. The second possibility is that the inner core can be unstable and, due to the quantum tunnelling effect, the spacetime can undergo a black hole-to-white hole transition (a bouncing Planck star).
\end{abstract}
\maketitle

\section{Introduction}
Using classical general relativity, in was shown by Oppenheimer and Snyder \cite{os}, that the ultimate fate of a spherically symmetric collapsing star must be a black hole. According to general relativity, classical black hole solutions have singularities arising during the gravitational collapse. In particular, Penrose showed that even deviations from spherical symmetry
cannot prevent space-time singularities from
arising \cite{penrose}.  On the other hand, Hawking \cite{hawkingbh}, used quantum field theory in strong gravitational field and found that there must be a thermal flux of particle production, known as Hawking radiation. This means that a static observer located far away from the black hole should detect temperature. Such temperature is very small and, as of today, it has not been measured.  It is widely believed that such spacetime singularities can be cured within a quantum theory of gravity. Regular black holes attracted a lot of attention (see different regular black hole solutions 
 \cite{Bardeen,Ayon-Beato:1998hmi,Haward,Frolov,Simpson:2018tsi,Jusufi:2022cfw}), including a recent review \cite{Sebastiani:2022wbz}, and possible constraints to the regular black holes with the Event Horizon Telescope image of Sagittarius A$^*$ and S2 star \cite{Vagnozzi:2022moj,s2}. Different concerns have been raised about the stability of regular black holes. Specifically, it was argued that regular black holes can be generically unstable because of the phenomenon known as the “mass inflation” which can destabilize the inner horizon and the role of Hawking radiation to cure this instability \cite{r1,r2}, and the problem with such a claim see \cite{r3}. 
 
 In the present paper, we shall peruse a different scenario, namely it was argued how ideas from T-duality can regularize the gravitational potential \cite{padma,Nicolini:2019irw,Smailagic,Nicolini:2022rlz} and how this can play an important role to resolve the black hole singularity \cite{Nicolini:2019irw}. Using T-duality, it is possible to show that the description of string theory below the length $l_s = \alpha' $  is  the same as its description above $l_s = \alpha' $. In this framework, the physics of  four  dimensions can be obtained by compactifying the other dimensions. For a single compact dimension with radius $R$, one can use the boundary conditions by writing
\begin{equation}
 X^4 (\tau, \sigma + 2 \pi) = X^4 (\tau, \sigma)
 + 2 \pi w R,
\end{equation}
where $w$ is known as the winding number. Furthermore, for the mass spectrum for such a system, we have
\begin{equation}
m^2 =\frac{1}{2\alpha'}\left(\, n^2 \frac{\alpha'}{R^2} +
w^2\frac{R^2}{\alpha'}\, \right)+ \dots,
\end{equation}
here $n$ is the known as the Kaluza-Klein excitation level.
The main idea behind T-duality is that the above spectrum  does not change if we exchange winding number $w$ and Kaluza-Klein excitation level $n$, namely we can write \cite{Nicolini:2019irw},
\begin{eqnarray}
w \to n, && R \to \alpha^{'2}/R.
\end{eqnarray}
On physical grounds, this also suggests that the description of string theory below a certain length is equivalent to its description above it \cite{padma}. Furthermore, by means of T-duality, one can show that Green's function is invariant under 
$  w \to n,  R \to \alpha^{'2}/R$. In particular, for the Green function in the momentum space, it was found the following \cite{padma,Nicolini:2019irw,Smailagic,Nicolini:2022rlz},
 \begin{equation}
  G(k) = -\frac{2 \pi R }{\sqrt{k^2}} K_1 ( 2 \pi R \sqrt{k^2}),
 \end{equation}
in which $K_1 (k)$ is a modified Bessel function of second kind. The zero point length ($l_0 = 2 \pi R$) is therefore produced by means of compactified extra dimension of radius $R$, and cannot be probed below this length.
Taking the limit $l_0 k^2 \to 0$,we get the standard relation for the Green's function is obtained, i.e., $G(k) = - k^{-2}$. In such a limit, the stringy effects are very small and can be totally neglected.
Using the modified Green's function it was also shown that the point-like source distribution is replaced by a smearing matter density. Specifically, using the regularized potential due to the zero-point length $l_0$, and by solving the Poisson equation one can obtain the energy density and the stress-energy tensor describing the smearing matter distribution \cite{Nicolini:2019irw}. The static and spherically
symmetric metric that solves the Einstein field equations with stringy effect is given by \cite{Nicolini:2019irw}
\begin{equation}\label{metric1}
ds^2=-\left(1-\frac{2Mr^2}{(r^2+l_0^2)^{3/2}}\right)dt^2+\frac{dr^2}{1-\frac{2Mr^2}{(r^2+l_0^2)^{3/2}}}+r^2 d\Omega^2,
\end{equation}
where $M$ denotes the Komar mass. This is a very important solution since it is a non-pertubative solution that describes a static and spherically symmetric black hole geometry.  For $M>3 \sqrt{3}\,l_0/4$, there exist two roots:
the inner and outer horizon, $r_-$ and $r_+$, respectively. We can also say that such a metric describes two possible phases of matter, the particle sector
($M<3 \sqrt{3}\,l_0/4$) and the black hole sector ($M>3 \sqrt{3}\,l_0/4$). For very large mass, the solution is effectively the Schwarzschild black hole.  Recently, such non-pertubative modification were used to find a charged black hole solution in T-duality \cite{k1}, regular black holes in 3 dimensions \cite{kr}, charged black holes in 4D Einstein-Gauss-Bonnet gravity \cite{kgbt}, entropic corrections to Friedmann equations \cite{k2}, regular black strings and torus-like black holes \cite{kt}. By considering the Hawking evaporation, it was argued that black remnants should occur due to stringy theoretical effects \cite{Pourhassan:2019luf}. 
In the present work, we would like to study in more details the gravitational collapse and the final stage of the collapse using such stringy corrections. 

This paper is outlined as follows. In Section II, we study the gravitational collapse of the interior of star. In Section III, we discuss the Planck star radius using an infalling observer point of view. In Section IV, we discuss the Hawking evaporation and the final Planck-size remnants. In Section V, we study the bouncing Planck star scenario. We comment on our
results in Section VI.

\section{Gravitational collapse and Planck stars in T-duality}
Let us start by considering a spherically-symmetric star composed by matter (dust) with vanishing pressure which undergoes a gravitational collapse.  In general, the
stress energy tensor of the collapsing matter is given by
\begin{equation}
T^{\mu \nu}=(\rho+p)u^{\mu} u^{\nu}+pg^{\mu \nu},
\end{equation}
in which  $\rho$ is the energy density, $u^{\mu}$ is the fluid 4-velocity, and $p$ is the pressure which in our case vanishes. For such a fluid we have to consider the energy conservation, i.e., $\nabla_{\alpha}T^{\alpha\beta} = 0$, along with the Einstein equations. For the interior region, having a spherically symmetry star, we can write in general  \cite{misner}
\begin{equation} 
ds_{\text{int}}^{2}=-e^{2\phi(r,\tau)} d\tau^{2}+ e^{\lambda(r,\tau)} dr^{2} + \mathcal{R}(r,\tau)^{2} d\Omega^{2},
\label{misnermetric}
\end{equation} 
in which $\mathcal{R}(r,\tau)$ is the area radius. In addition, we take $\phi'(r,\tau) =0$, which follows from the Einstein equations for the case of homogeneous dust \cite{misner}. To study the gravitational collapse we are going to use the well known Tolman-Bondi spacetime by introducing following function \cite{Hajicek:2001yd,Liu:2014kra,Bambi:2013caa,Mersini-Houghton:2014zka,lec,Kiefer:2019csi}
\begin{eqnarray}
e^{\lambda(r,\tau)}=\frac{\left[\mathcal{R}(r,\tau)\right]_{,r}^2}{1-\mathcal{K}(r)},
\end{eqnarray}
which leads to
\begin{equation} 
ds_{\text{int}}^{2}=-d\tau^{2}+ \frac{\left[\mathcal{R}(r,\tau)\right]_{,r}^2}{1-\mathcal{K}(r)} dr^{2} + \mathcal{R}(r,\tau)^{2} d\Omega^{2}.
\end{equation} 

For the exterior metric, we shall use the modified vacuum solution due to the stringy effects given by Eq. (5) 
 and rewritten as
\begin{equation} 
ds_{\text{ext}}^{2}=-\left(1-\frac{2Mr^2 }{(r^2+l_0^2)^{3/2}}\right)dt^{2}+\frac{dr^{2}}{1-\frac{2Mr^2 }{(r^2+l_0^2)^{3/2}}}+r^{2}d\Omega^{2}.
\end{equation} 

At this point, we will utilize the Misner-Sharp mass
function which is defined by using the area radius, which at fixed $\mathcal{R}$, reads
\begin{eqnarray}
g^{\mu \nu}(\nabla_{\mu}\mathcal{R})(\nabla_{\nu} \mathcal{R})=1-\frac{2M\mathcal{R}^2 }{(\mathcal{R}^2+l_0^2)^{3/2}},
\end{eqnarray}
from this equation it follows that 
\begin{eqnarray}
\left[\mathcal{R}(r,\tau)\right]_{,\tau}^2=\frac{2M\mathcal{R}^2 }{(\mathcal{R}^2+l_0^2)^{3/2}}-\mathcal{K}(r).
\end{eqnarray}

A very important result that follows from the Tolman-Bondi spacetime is that one can obtain the
Friedmann’s equations as a special case. To see this, we need to introduce the following relations
\begin{eqnarray}
\mathcal{R}(r,\tau)=a(\tau)r\,\,\,\,\text{and}\,\,\,\,\,\mathcal{K}(r)=k\,r^2.
\end{eqnarray}
It can be easily seen how the FRW universe metric is obtained
\begin{equation}
ds^{2} = - d\tau^{2} + a(\tau)^{2} \left(\frac{dr^{2}}{1 - k\, r^{2}} + r^{2} d\Omega^{2} \right).
\end{equation}
Here $k$ denotes the
curvature of space with $k = 0, 1, -1$ corresponding to flat,
closed, and open universes, respectively. Using this equivalence we can model and study the interior spherically symmetric homogeneous stars. Using Eqs. (12) and (13) we can find
\begin{equation}
\left(  \frac{\dot{a}}{a}\right)^2+\frac{k}{a^2}=\frac{8 \pi \rho  }{3}\left(1+\frac{l_0^2}{\mathcal{R}^2}\right)^{-3/2}.
\end{equation}

It is important to note here that we shall neglect the backreaction effect of the pre-Hawking radiation during the gravitational collapse. In particular, it has been shown that such an effect is small and cannot prevent the formation of apparent horizon (see for example \cite{Chen:2017pkl}).
The dynamical apparent
horizon, a marginally trapped surface with vanishing expansion, is
determined by the relation
\begin{eqnarray}
h^{\mu
\nu}\left(\partial_{\mu}\mathcal{R}\right)\left(\partial_{\nu}\mathcal{R}\right)=0,
\end{eqnarray}
where the two dimensional metric reads
\begin{equation}
    h_{\mu \nu}=\textrm{diag}(-1, \frac{a^2}{1-kr^2}).
\end{equation}
It is a simple calculation to find out the relation for
the apparent horizon radius of the FRW universe 
\begin{equation}
\label{radius}
 \mathcal{R}=a\,r=\frac{1}{\sqrt{\left(\frac{\dot{a}}{a}\right)^2+\frac{k}{a^2}}}.
\end{equation}
We are going to simplify the work, since $l_0$ is a very small number, we can consider a series expansion around $l_0$ via
\begin{eqnarray}
\left[1+\frac{l_0^2}{r^2 a^2}\right]^{-3/2}=1-\frac{3}{2}\frac{l_0^2}{r^2 a^2}+...
\end{eqnarray}
then using the Friedmann's equation (15) we obtain in leading order terms
\begin{eqnarray}
\left(  \frac{\dot{a}}{a}\right)^2+\frac{k}{a^2}=\frac{8 \pi \rho  }{3}\left[1-\frac{3 l_0^2}{2} \left(  \left(\frac{\dot{a}}{a}\right)^2+\frac{k}{a^2}\right)\right].
\end{eqnarray}

The last equation can be further written as
\begin{equation}\label{Fried2}
\left( \frac{\dot{a}}{a}\right)^2+\frac{k}{a^2} =\frac{8\pi \rho }{3}\left[1-\Gamma \rho\right],
\end{equation}
where $\Gamma$ is a constant defined as
\begin{equation}\label{Gamma}
\Gamma\equiv\frac{4l_0^2\pi }{3}.
\end{equation}
This result is nothing but the corrected Fredmann equation reported recently in Ref. \cite{k2} using a different approach (Verlinde's entropic force scenario). In fact, it coincides with \cite{k2} by taking $\omega=0$ (dust) \footnote{In fact, to get the precise correspondence with \cite{k2}, in the second term of the r.h.s of Eq. (21) we make the replacement $\rho \to \rho/3$. Compared to the fully relativistic form \cite{k2}, the factor three in the corrected term is probably related to the fact that here $\rho$ is interpreted as the averaged mass density}. It is quite remarkable that we found a bridge between two different and competing directions in quantum gravity.  In one hand, by considering string T-duality effects we found modified Friedmann equations (21) which coincides with the conclusions obtained from the loop quantum gravity approach  \cite{LQG}
\begin{equation}
\left( \frac{\dot{a}}{a}\right)^2+\frac{k}{a^2} =\frac{8\pi \rho }{3}\left[1-\frac{\rho}{\rho_c}\right],
\end{equation}
where $\rho_c$ is the critical energy density
\begin{equation}
\rho_c\equiv \frac{3}{8 \pi \gamma^2 \lambda^2 },
\end{equation}
where $\lambda \sim 5.2 l_{Pl}^2$ \cite{LQG} is area gap  that sets the discreteness scale of loop quantum gravity and $\gamma$ is the Immirzi parameter. The correspondence is achieved by identifying $\Gamma=\rho_c^{-1}$. A direct computation yields $\gamma= 0.310086 \,l_0/l_{Pl}$. Using $l_0=2^{3/4}/3^{3/4} l_{Pl}=0.73778 l_{Pl}$ \cite{Nicolini:2019irw}, we get $\gamma=0.2287783$, which is in perfect agreement with the value proposed in loop quantum gravity  $\gamma=0.2375$. 
Once the gravitational collapse takes place, we can now use the modified Friedmann equations and explore the possibility that the collapse stops at some point due to the stringy corrections. To do so, we have to use the condition
\begin{equation}
\dot{a}=0|_{(a=a_{\text{min}},\rho=\rho_{\text{crit.}})},
\end{equation} 
along with $k=1$. From this condition, we can get the critical density, in fact we obtain 
two branches of solution for the critical
density
\begin{equation}
\rho_{\text{crit.}}=\frac{1}{2 \Gamma}\left(1\pm \sqrt{1-\frac{3 \Gamma}{2 \pi a_{\text{min}}^2}}\right).
\end{equation}
From this result it follows that  
\begin{eqnarray}
1-\frac{3 \Gamma}{2  \pi a_{\text{min}}^2}
\geq 0,
\end{eqnarray}
which basically allows us to find the minimal quantity for the scale
factor
\begin{equation}
a_{\text{min}}=\sqrt{\frac{3 \Gamma }{2  \pi}}=\sqrt{2}\, l_0.
\end{equation}
Again, this is in perfect agreement with what has been found in Ref. \cite{k2}. Such a critical density is thus inversely proportional to the minimal length
\begin{equation}
\rho_{\text{crit.}} \sim \frac{1}{l_0^2}.
\end{equation}

The above arguments show that during the  gravitational collapsing phase, the singularity is never reached and the interior solution of the black hole (black hole core) is a kind of very dense star. This possibility that a very dense star or Planck star exists inside the black hole was proposed in Ref. \cite{Rovelli:2014cta}. The radius of such a star was conjectured to be proportional to the collapsed mass \cite{Rovelli:2014cta}. It is very interesting, as we shall see, such Planck stars hidden inside the stringy corrected regular black holes can naturally appear in our analyses. In what follows, we are going to compute the radius of such a star. First, we need to rewrite the FRW metric in a simple form. Let us define the proper time $\tau$ using
\begin{eqnarray}
\tau &=& \int a(\eta)d \eta,
\end{eqnarray}
where $\eta$ is the conformal time, along with radial coordinate defined as
\begin{eqnarray}
r(\tau)&=& a(\tau)\sin \chi.
\end{eqnarray}
From these equations we obtain the FRW metric as
\begin{eqnarray}\notag
ds_{\text{int}}^{2}&=&-d\tau^{2}+a^{2}(\tau)\left[d\chi^{2}+\sin^{2}\chi dr^{2}\right]\\
&=& a^{2}(\eta)\left[-d\eta^{2}+d\chi^{2}+\sin^{2}\chi dr^{2}\right].
\end{eqnarray}
Choosing a surface $\Sigma$, with fixed $\chi=\chi_0$, by matching the metrics, we can obtain the first equation
\begin{equation}
    \mathcal{R}(\tau)=a(\tau) \sin\chi_0,
\end{equation}
along with the second equation 
\begin{equation}
-\left(1-\frac{2M\mathcal{R}^2 }{(\mathcal{R}^2+l_0^2)^{3/2}}\right)\left(\frac{dt}{d\tau}\right)^2+\frac{\left(\frac{d\mathcal{R}}{d\tau}\right)^2}{1-\frac{2M\mathcal{R}^2 }{(\mathcal{R}^2+l_0^2)^{3/2}}}=-1.
\end{equation}
From the last equation it is not difficult to show that
\begin{eqnarray}
\frac{dt}{d\tau}=\pm \frac{\sqrt{\dot{\mathcal{R}}^2+1-\frac{2M\mathcal{R}^2 }{(\mathcal{R}^2+l_0^2)^{3/2}}}}{1-\frac{2M\mathcal{R}^2 }{(\mathcal{R}^2+l_0^2)^{3/2}}}.
\end{eqnarray}
Although we use a rather simple and idealized model of collapse, it highlights the main features of the interior dynamics of interior of the star. Using the matching procedure of the interior and exterior metrics at the surface of the star it is possible to study the motion of the star's surface. In what follows we shall show some important results; first, we are going to approximate the last equation as 
\begin{eqnarray}
dt\simeq \pm \frac{d \mathcal{R}}{1-\frac{2M\mathcal{R}^2 }{(\mathcal{R}^2+l_0^2)^{3/2}}},
\end{eqnarray}
where plus/minus sign corresponds to the case of expansion or collapse. Since we are interested in the collapse, we chose the minus sign ($\mathcal{R}$ decreases with time) and by doing further simplifications we get 
\begin{equation}
\mathcal{R}(t)\simeq 2M-\exp\left(-\frac{4tM+8M^2+8M^2\Xi+3l_0^2 \Xi}{8M^2+3l_0^2}\right),
\end{equation}
where
\begin{eqnarray}
\Xi=\text{LabertW}\left(- \frac{4M e^{-\frac{4M(t+2M)}{8M^2+3l_0^2}}}{8M^2+3l_0^2}  \right).
\end{eqnarray}
We see that from the point of view of the outside observer, it takes infinite amount of time $t \to \infty$ to see the formation of the black hole horizon $\mathcal{R}\to 2 M$. The whole process is viewed in “very slow motion”.  However, from the point of view from the inside, it takes a finite proper time for particles to reach the minimal distance. In the Oppenheimer-Snyder model ~\cite{os}, the surface of a gravitationally collapsing spherically symmetric star made up of dust with radius $R_s$, can be obtained via Eq. (33). At this point, let us define the following constant quantity
\begin{eqnarray}
a_0 \equiv \frac{8 \pi \rho a^3}{3}|_{(\tau =0)}={\text{const.}}    
\end{eqnarray}
where $a_0=a (\tau =0)$ is the scale factor in the initial moment of collapse. We can see that this quantity is constant simply by taking $\rho=\rho_0 a^{-3}$, for the dust matter. At the initial time we also have $\tau=\eta=0$, along with radius of the star $R_{s}(0)=R_{0}$. Furthermore one can show that 
\begin{eqnarray}
a_0= \sqrt{\frac{R_{0}^{3}}{2M}},\,\,\,\sin\chi_{0}=\sqrt{\frac{2M}{R_{0}}}.
\end{eqnarray}
From the corrected Fredmann's equation [setting $k=1$], we obtain
\begin{eqnarray}
\dot{a}(\tau)+1=\frac{a_0}{a(\tau)}\left(1-\frac{a_0 l_0^2}{2 a(\tau)^3}  \right)
\end{eqnarray}
or, in terms of $\eta$, we get
\begin{eqnarray}
\dot{a}(\eta)+a(\eta)^2=a_0 a(\eta)\left(1-\frac{a_0 l_0^2}{2 a(\eta)^3}  \right).
\end{eqnarray}

Solving the last two equation exactly is not an easy task. 
One simple guess is to try and generalize the parametric form $a(\eta)=a_0(1+\cos \eta)/2$ which is a solution when $l_0=0$, then by using the following equation
\begin{eqnarray}
a(\eta)=\frac{a_0}{2}\left( 1+\xi(\eta)\right)
\end{eqnarray}
we get
\begin{equation}
\xi(\eta)=\eta \pm \int \frac{(1+ a(\eta))a_0 d a(\eta)}{\sqrt{(1-a(\eta))(1+a(\eta))^3-8 l_0^2}}+C.
\end{equation}
There are two branches in this solution which can describe the contraction and expansion, respectively.  Again, finding an exact solution in closed form is outside the scope of the present work. Since the mass is conserved during the gravitational collapse (having in mind the Hawking radiation is very small) we must also have 
\begin{eqnarray}
a(\tau_{\text{max}}) \equiv \frac{8 \pi \rho a^3}{3}|_{(\tau_{\text{max}})}={\text{const.}}  
\end{eqnarray}
once the Planck star is formed. At the surface of the star when the gravitational collapse stops, we also have 
\begin{eqnarray}
a(\tau_{\text{max}})= \sqrt{\frac{R_{\tau_{\text{max}}}^{3}}{2M}},\,\,\,\sin\chi_{\tau_{\text{max}}}=\sqrt{\frac{2M}{R_{\tau_{\text{max}}}}},
\end{eqnarray}
note here that $a(\tau_{\text{max}})=a_{\min}=\sqrt{2}\, l_0$. This means that we can obtain the proportionality
\begin{eqnarray}
\rho_0 a_{0}^2=\rho (\tau_{\text{max}}) a_{\min}^2,
\end{eqnarray}
where we can identify $\rho (\tau_{\text{max}})=\rho_{\text{crit}}$. Put in other words, during the gravitational collapse, the scale factor decreases, but the density per unit volume increases. Another way of stating this result is to say the mass of the collapsing matter is constant
\begin{eqnarray}
\rho_0 R_{0}^3=\rho_{\text{crit}} R_{\tau_{\text{max}}}^2.
\end{eqnarray}

When the gravitational collapse stops, we can find  the radius of the Planck star using $R_{s}|_{\tau_{max},a_{\min}}= a(\tau_{\text{max}}) \sin \chi_{\tau_{\text{max}}}$, namely we get
\begin{equation}
 R_{s}|_{\tau_{{\text{max}}},a_{{\text{min}}}}\sim a_{{\text{min}}} (2M)^{1/2} R_{\tau_{max}}^{-1/2}.
\end{equation}

Using $R_s=R_{s}|_{\tau_{\text{max}},a_{\text{min}}}=R_{\tau_{\text{\text{max}}}}$, we estimate the radius of Planck star  as follows
\begin{eqnarray}
R_{s}\sim 2^{2/3}\,M^{1/3}\, l_0^{2/3}.
\end{eqnarray}
   The above value for the radius is in good agreement with Ref. \cite{Rovelli:2014cta}, having set $n=1/3$. The phenomenological aspects of Planck stars have been studied in Refs. \cite{pls1,pls2,pls3,pls4}. For a stellar mass black hole with mass $M \sim 10 \times M_{\text{sun}}$ and $l_0 \sim 10^{-34}$ m we can obtain the radius of the Planck star [by restoring the constants $G$ and $c$] $R_{s}\simeq [G Ml_0^2/c^2]^{1/3} \sim  10^{-22}$ m. Although a small value, this shows that the radius of such a star is many order of magnitudes greater compared to $l_0$. Such a star is hidden inside the event horizon of the black hole with the geometry described by the metric (5). 

\section{A free falling observer and Planck star radius}
Let us now study the whole process as seen from a free falling observer. To do this, we can use the Painlev\'{e}--Gullstrand coordinates through the  definition of a new time coordinate as 
\begin{equation}\label{Painleve time}
dt_{p}=dt+\frac{\sqrt{1-f(r)}}{f(r)}dr
\end{equation}
for some arbitary function $f(r)$, along with the new metric
\begin{equation}\label{painleve metric}
ds^2=-f(r)dt_p^2+2\sqrt{1-f(r)} dt_{p} dr+dr^2+r^2 d\Omega^2\,.
\end{equation}
We see that there is no coordinate singularity at the horizon. The time coordinate of the Painlev\'{e}--Gullstrand metric is the same as the proper time of a freely-falling observer who starts from infinity at zero velocity. We denote the Painlev\'{e}--Gullstrand coordinates as $(t_{p},r_{p})$ and the Schwarzschild coordinates as $(t,r)$. One can use the Jacobian to relate these coordinates given by \cite{Dey}
\[
\frac{\partial (t_{p},r_p)}{\partial(t,r)}=
\left({\begin{array}{cc}
\frac{{\partial t_{p}}}{{\partial t}}&\frac{{\partial t_{p}}}{{\partial r}}\\
\frac{{\partial r_p}}{{\partial t}}&\frac{{\partial r_p}}{{\partial r}}\\
\end{array}}\right)
=\left({\begin{array}{cc}
1& \frac{\sqrt{1-f(r)}}{f(r)}\\
0&1\\
\end{array}}\right)\,,
\]
along with the inverse of the transformation matrix 
\[
\frac{\partial (t,r)}{\partial(t_{p},r_p)}=
\left({\begin{array}{cc}
\frac{{\partial t}}{{\partial t_{p}}}&\frac{{\partial t}}{{\partial r_p}}\\
\frac{{\partial r}}{{\partial t_{p}}}&\frac{{\partial r}}{{\partial r_p}}\\
\end{array}}\right)
=\left({\begin{array}{cc}
1&-\frac{\sqrt{1-f(r)}}{f(r)}\\
0&1\\
\end{array}}\right)\,.
\]
From the point of view of a static observer located far away from the black hole, the total energy momentum is the sum of the energy momentum of the black hole core or the Planck star energy density and the renormalized stress energy tensor is we add the effect of Hawking radiation. For example, one can choose the Unruh vacuum stat [see, \cite{Dey}]. Hence we can write 
\begin{eqnarray}
T^{\text{tot.}} _{\mu\nu}=T^{\text{Core}} _{\mu\nu}+T^{\text{RSET}} _{\mu\nu}
\end{eqnarray}
For a freely-falling observer we can write the components in Painlev\'{e}--Gullstrand coordinates by means of a coordinate transformation 
 \begin{equation}
 T^{\text{GP}}_{\alpha\beta}=\frac{\partial x^{\mu}}{\partial x^{\alpha}}\frac{\partial x^{\nu}}{\partial x^{\beta}} T^{\text{tot}} _{\mu\nu}\,.
 \end{equation}
 As we did in the last section, we shall neglect here too the Hawking radiation effect as perceived by a freely-falling observer, and focus only on $T^{\text{Core}}$. For simplicity we work in $1+1$ dimensions, this yields
 \begin{eqnarray}\label{Ttptp}
 T_{{t_{p}}{t_{p}}}&=& f(r) \rho(r)\\
 T_{{t_{p}}{r_p}}&=& - \sqrt{1-f(r)} \rho(r),\\
T_{{r_p}{r_p}}&=&- \rho,
 \end{eqnarray}
with the components of the energy-momentum for the black hole core in Schwarzschild coordinates given by
\begin{eqnarray}
{{T^{\text{Core}}}^{\mu}}_{\nu}=\left(-\rho,P_r\right)
\end{eqnarray}

 For a freely-falling observer, the velocity in Painlev\'{e}--Gullstrand coordinates is given by
 \begin{equation}\label{free falling velocity}
 V^{a}=\left(1,-\sqrt{1-f(r)}\right).
 \end{equation}
 Using this velocity, we find that the energy density as measured by such an observer is given by
 \begin{equation}\label{energy density}
{\rho}_{\text{GP}}=T_{ab} V^{a} V^{b}=\rho\,.
 \end{equation}
In other words, the energy density stays invariant quantity. At this point, we use the condition
\begin{eqnarray}
V^a V_a=-1\,\,\,\Longrightarrow f(r)'|_{r=r_{\text{min}}}=0,
\end{eqnarray}
and after solving this equation we get the minimal value at
\begin{eqnarray}
r_{\text{min}}=\sqrt{2}\, l_0.
\end{eqnarray}
This is in perfect agreement with the minimal scale factor found in the last section. We can now compute the total time measured by such a free falling observer using
\begin{eqnarray}
\int_0^{T} dt_p= - \int_{r_+}^{r_{min}} \frac{dr}{\sqrt{1-f(r)}}
\end{eqnarray}
where we approximate $r_+\simeq 2M$. After solving this integral we obtain 
\begin{eqnarray}
T &\simeq& \frac{4}{3}M+\frac{2^{1/4} l_0^{2/3}}{12 \sqrt{M} }-\frac{3 l_0^2}{4M }+....
\end{eqnarray}

The proper time of a particle is therefore finite. We can show that the time for light reaching the minimal distance, say from event horizon is also finite. In this case, one can use $ds^2$, to find the radial equations, then using the integrating the equation we obtain
\begin{eqnarray}
\int_0^{T} dt_p= - \int_{r_{max}}^{r_{min}} \frac{dr}{1+\sqrt{1-f(r)}}
\end{eqnarray}
where again we can use the approximation $r_{max}\simeq 2M$. After solving this integral we obtain a finite amount of time
\begin{eqnarray}\notag
T &\simeq& 2M \ln M+6 M \ln 2 -2 M+ 2 \sqrt{ 2 M } \sqrt{l_0\sqrt{2} }\\
&-& \sqrt{2} l_0 -4 M \ln(\sqrt{ 2 M } \sqrt{l_0 \sqrt{2}})
\end{eqnarray}

Let us now use Eq. (51) to find the time $t$ measured by an observer located far away from the black hole 
\begin{eqnarray}
t=T-\int\frac{\sqrt{1-f(r)}}{f(r)}dr
\end{eqnarray}
which yields
\begin{equation}
t\simeq T+2 \sqrt{2M}\left(\sqrt{2M}\arctan^{-1} \left(\sqrt{\frac{r}{2M}}\right)-\sqrt{r}   \right)+C
\end{equation}
where $C$ is an integration constant. In the limit $r\to 2M$, we obtain $t\to \infty$, meaning that from this observer point of view, it takes an infinite amount of time to see the collapsing of matter. Due to the quantum gravitational effect, or the zero point length effect, we found that the particles never reach the singularity, but this also implies the existence of Plank stars.  This can be seen from Einstein field equations and using $\rho(r)=\rho_{\text{crit.}}$, we must have
\begin{eqnarray}
 \mathbf{R} \sim 8 \pi \rho_{\text{crit.}}(r) \sim \frac{3}{l_0^2}.
\end{eqnarray}

This shows that there is no singularity in the expression for the Ricci scalar, provided $l_0>0$. One can calculate one more scalar invariant, known as the Kretschmann scalar given by the following result $\mathbf{K}\sim l_0^{-4}.$ 
To estimate the radius of the Planck star, here we recall that the Ricci scalar ($\mathbf{R}$) for the above black hole given is found 
\begin{equation}
    \mathbf{R}=\frac{2 M \left(2 r^4-11 l_0^2r^2+2 l_0^2 \right)}{(l_0^2+r^2)^{7/2}}.
\end{equation}

From these two equations we obtain
\begin{eqnarray}
\frac{2 M \left(2 r^4-11 l_0^2r^2+2 l_0^2 \right)}{(l_0^2+r^2)^{7/2}}-\frac{3}{l_0^2}=0,
\end{eqnarray}
considering a series expansion around $l_0$, and by setting the radial coordinate to be the Planck star radius [we call it $r=R_s$] we get
\begin{eqnarray}
\frac{4M}{R_s^3}-\frac{3}{l_0^2}=0.
\end{eqnarray}

Solving for the $R_s$ we obtain 
\begin{eqnarray}
R_s \sim 2^{2/3}3^{-1/3} M^{1/3} l_0^{2/3},
\end{eqnarray}
which is in perfect agreement with Eq. (50) found in the last section in leading order terms.

\section{Planck-size remnants}
In this last section, we would like to speculate about the final state of the Planck star hidden inside the black hole. Assuming that black hole has been formed along with a Planck star inside it, due to the presence of the horizon, we can now take into the account the Hawking radiation and its back
reaction effect. Viewed from the outside region, we have the outer horizon \cite{Nicolini:2019irw}
\begin{eqnarray}
r_+\simeq 2M-\frac{3l_0^2}{4M},
\end{eqnarray}
and the inner horizon 
\begin{eqnarray}
r_{-}\simeq \frac{l_0}{\sqrt{2}}\left( \frac{l_0}{M}\right)^{1/2},
\end{eqnarray}
respectively. The Hawking radiation is computed via \cite{Nicolini:2019irw}
\begin{eqnarray}
T_H=\frac{f'(r)}{4\pi}|_{r=r_+}=\frac{1}{4 \pi r_+}\left(1- \frac{3 l_0^2}{l_0^2+r_+^2} \right).
\end{eqnarray}
Due to the backreaction effect of the Hawking  evaporation the mass of the black hole decreases $M(t)$, this means that we have a slowly shrinking outer horizon, but in the same time the inner horizon increases [as can be seen from Eqs. (74)-(75)]. For instance, we can compute the evaporation time viewed from the outside using
\begin{equation}
-\frac{dM(t)}{dt} \sim A\,\sigma\, T_H^4 
\end{equation}
where $A=4 \pi r_+^2$ is the area of the black hole horizon and $\sigma$ is the Stefan–Boltzmann constant. For the evaporation time it is not difficult to show that
\begin{eqnarray}
t_{\text{evapo.}}\simeq \mathcal{A}\left(M^3-M_{\text{ext}}^3\right)+l_0^2 \mathcal{B} \left(M-M_{\text{ext}} \right).
\end{eqnarray}
where $\mathcal{A}$ and $\mathcal{B}$ are two constants of proportionality. The stringy effects are small and the evaporation time will be very long. It was shown that for some extremal configuration with $M=M^{\text{ext}}=3 \sqrt{3} l_0/4$ the outer and inner horizon coincide $r_-=r_+=\sqrt{2} l_0$  (see, for details  \cite{Nicolini:2019irw}). This is interesting since it coincides exactly with the minimal scale factor obtained in the present work. There is a significant difference compare to the classical Schwarzschild black hole case, namely instead of getting increasingly hotter and eventually with a
final explosion, due to the stringy effect, here it cools down and eventually vanishes $(T_H=0)$ at the
extremal configuration. This offers a possibility that the final state - which is a result of a very long time, to be 
stable remnant with
\begin{eqnarray}
R_s^{\text{ext}}\sim r_-=r_+=\sqrt{2}\, l_0,
\end{eqnarray}
 This small mass of the remnant is nothing but a particle, and it has been speculated to be a candidate for the dark matter. 

\section{Bouncing Planck star: Black hole to white hole transition}
There is another possibility, perhaps a more interesting one, in which, the Planck star bounces instead of decreasing it's radius. This is due to the fact that that the inner core (Planck star) solution may not be a stable state after all. Mathematically, the bouncing at the critical point can be stated using the conditions: $a_{\text{min}}>0$, $H|_{a=a_{\text{min}}}=0$, along with the condition $\ddot{a}|_{a=a_{\text{min}}}>0$.  This shows that there is a great level of similarity between the physics that describes the cosmic bounce and the possible bounce inside black holes. One can use the second modified Friedmann equation that describes the dynamical evolution reported in Ref. \cite{k2}
\begin{equation}\label{addot0}
\frac{\ddot{a}}{a}=- \left(\frac{4 \pi  }{3}\right)(\rho_{\text{crit}}+3p_{\text{crit}})\left[1-\frac{3}{2} \frac{l_0^2}{a_{\text{min}}^2(\omega)}+... \right],
\end{equation}
with the minimal scale factor given by \cite{k2} 
\begin{eqnarray}
a_{\text{min}}(\omega)=\sqrt{2}\,\sqrt{\frac{1+3\omega}{1+ \omega}}\, l_0.
\end{eqnarray}
We see that in general, if we have matter with non-vanishing pressure then the scale factor can be a function of $\omega$. 
Imposing the condition $a_{\text{min}}>0$, we get the interval $\omega \in (-\infty, -1) \cup (-1/3, \infty)$. On the other hand, if we use the equation of state via $p_{\text{crit}}=\omega \rho_{\text{crit}}$, along with $\rho_{\text{crit}}=1/(2\Gamma)$, we obtain 
\begin{equation}\label{addot0}
\frac{\ddot{a}}{a}=-\frac{1}{l_0^2}\frac{1+9 \omega}{8},
\end{equation}
which is further rewritten as 
\begin{eqnarray}
 \ddot{a}(\tau)-\zeta^2\, a(\tau)=0,
\end{eqnarray}
with \begin{eqnarray}
\zeta^2=-\frac{1}{l_0^2}\frac{1+9 \omega}{8}>0,
\end{eqnarray}
provided  $\omega<-1/9$. But we must also have in mind that $a_{\text{min}}>0$, therefore, we are left with the allowed interval $-1/3 < \omega <-1/9$. The general solution in this interval is given by
\begin{eqnarray}
a(\tau)=B_1 \exp{\left(\zeta \tau\right)}+B_2 \exp{\left(-\zeta \tau\right)},
\end{eqnarray}
where we can take the interval $a_{\text{min}}\leq  a(\tau) \leq a_{\text{max}} $. At the initial moment $\tau=0$, one has $a(\tau=0)=a_{\text{min}}=B_1$, hence we can fix the constant $B_2=0$, which yields $a(\tau)=a_{\text{min}}\exp(\zeta \tau)$. The interior metric now reads 
\begin{eqnarray}
ds_{in}^{2}&=&-d\tau^{2}+a^2_{\text{min}}e^{2\zeta \tau}\left[d\chi^{2}+\sin^{2}\chi dr^{2}\right]
\end{eqnarray}

As was argued in \cite{k2}, this metric can describe the bouncing universe.  For reasons we elaborated above, we need a special form of matter with a specific interval for EoS parameter $\omega$ in order to justify the bouncing effect. Coming back to our case, where we studied the collapsing of matter (dust) with zero pressure i.e., $p_{\text{crit}}=0$, along with $\omega=0$, this means that the above bouncing condition is not satisfied.  At this point a natural question arises: even if we have a collapsing dust which clearly does not satisfy the above bouncing condition, can we still say that the final state of the internal core of the black hole will be eternally stable? Of course, we don't know the answer to this question, but from a quantum mechanical point of view, we may speculate that the bouncing effect can be also a consequence of the black hole-to white hole transition (BHWH). In other words, instead of the bouncing condition given by Eq. (84) which is classical effect, we can have a purely quantum mechanical bounce due to the quantum tunneling effect.  The idea behind the BHWH transition is not new, for example in \cite{carlos}, authors have tried to compute the probability amplitude between two configurations, say $h_{-}$ and $h_{+}$, with the corresponding hypersurfaces $\Sigma_{-}$ and $\Sigma_{+}$.  In particular, the probability amplitude for the BHWH transition  can be computed from \cite{carlos}
\begin{equation}
\mathcal{P}_{BH \to WH}(M, \Delta_0)=\int_0^{\Delta_0}|\braket{WH|BH}_{M,\Delta^{\prime}_0}|^2 \,\mathrm{d}\Delta^{\prime}_0,
\end{equation}
where $\Delta_0$ is a parameter measuring the width of the interpolating region. Furthermore, it was estimated for the BHWH transition probability a exponential decay law \cite{carlos}
\begin{equation}
\mathcal{P}_{BH \to WH}(M, \Delta_0)\simeq 1-e^{-M \Delta_0},
\end{equation}
with a mean lifetime  $\tau \leq 1/2M$. There are other other arguments about the black hole-to-white hole transition. For instance, the probability increases with time if we take into account the Hawking radiation (see, \cite{pls2,Bianchi:2018mml}). This can be explained from the fact that as the mass of the Planck star decreases with time and the bouncing mass will be smaller compared to the initial Planck mass, then in accordance with the semiclassical standard tunnelling factor $\sim e^{-S_E/\hbar}$  \cite{pls2,Bianchi:2018mml},
 the probability for the black hole-to-white hole transition increases as the mass decreases. Here we note that $S_E$ is the Euclidian action, where $S_E=M^2$. The tunnelling probability per unit time can also be written [here we restore $\hbar$ for a moment] 
$\mathcal{P}_{BH \to WH} \sim e^{-M^2/\hbar}/M$ \cite{pls2,Bianchi:2018mml}.
Now let us consider again the Painlev\'{e}--Gullstrand coordinates that relate the time measured by an outside observer and the time measured by an outgoing observer given by
\begin{equation}\label{Painleve time}
dt_{p}=dt-\frac{\sqrt{1-f(r)}}{f(r)}dr,
\end{equation}
along with the white hole metric
\begin{equation}\label{painleve metric}
ds^2_{WH}=-f(r)dt_p^2-2\sqrt{1-f(r)} dt_{p} dr+dr^2+r^2 d\Omega^2\,.
\end{equation}
  Due to the bouncing effect, the black hole becomes essentially a white hole with an explicit time-reversal symmetry. In that sense, a white hole is a solution in general relativity, with a spacetime region to which cannot be entered from the outside.  From the point of view of an outside observer measuring in Schwarzschild coordinates, the time-reversed solution or the white hole geometry is the same as black hole.  As we saw, it takes a finite proper time to form a Planck star from the gravitational collapse and yet from the point of view of outside observer, due to the strong redshift effect, the gravitational collapse appears 'frozen' in time due to the formation of the horizon. The same can be shown for the bouncing process. An outside observer sees the collapse/bouncing in “very slow motion”, and the entire process takes a long time. To see this, let us consider a white hole region with a time-reversed solution, i.e., $t \to -t$ in Eq. (89), then the time measured outside the white hole is given by
\begin{eqnarray}
t=\int {dt_{p}}+\int \frac{\sqrt{1-f(r)}}{f(r)}dr.
\end{eqnarray}
The first terms is finite proper time measured by the outgoing observer and can be computed via 
\begin{eqnarray}
T=\int_{0}^T {dt_{p}}=\int_{r_{\text{min}}}^{r_{\text{max}}} \frac{dr}{\sqrt{1-f(r)}},
\end{eqnarray}
and the result is similar to Eq. (64). For the time measured by the outside observer we get
\begin{equation}
t\simeq T+2 \sqrt{2M}\left(\sqrt{2M}\arctan^{-1} \left(\sqrt{\frac{r}{2M}}\right)-\sqrt{r}   \right)+C,
\end{equation}
meaning that a particle to reach the event horizon $r \sim 2M$, we need $t \to \infty $.  Put in other words, the bouncing effect of the star appears in “very slow motion” when observed from the outside.  We can basically deduce the same conclusion using the matching of the interior and exterior metrics. To do such a computation we need of course the explicit form of the scale factor. Let us take just for fun the exponential law i.e., $a(\tau) \sim \exp(\zeta \tau)$, then we get 
\begin{equation}
    \mathcal{R}(\tau)= a_{\text{min}}\exp(\zeta \tau) \sqrt{\frac{2M}{R_s}},
\end{equation}
where $\zeta=C/l_0$, here $C$ is some constant. In the initial time of expansion $\tau=0$, we must have $ \mathcal{R}(\tau=0)=R_s$, i.e. $ \mathcal{R}$ should coincide with the radius of the Planck star. Now assuming that during the expansion we reach the classical horizon radius with $\mathcal{R}\to 2 M$, we get the total proper time 
\begin{eqnarray}
\tau \simeq \frac{l_0}{2 C}\ln\left(\frac{2MR_s}{a_{\text{min}}^2}  \right).
\end{eqnarray}

Since we have an expansion in this case, we need to take the plus sign in the right hand side of Eq. (36), along with the time-reversed condition $t \to -t$. In doing so, we get
\begin{eqnarray}
dt \simeq \frac{d \mathcal{R}}{\frac{2M\mathcal{R}^2 }{(\mathcal{R}^2+l_0^2)^{3/2}}-1}\simeq  \frac{d\mathcal{R}}{\frac{2M}{\mathcal{R}}-1}.
\end{eqnarray}

Solving this equation for the time leads to the following result
\begin{equation}
t=-2M\ln\left(M\sqrt{2}-e^{\frac{C \tau}{l_0}}a_{\text{min}}\sqrt{\frac{M}{R_s}}\right)-e^{\frac{C \tau}{l_0}}a_{\text{min}} \sqrt{\frac{2M}{R_s}}.
\end{equation}
If we replace the expression for the proper time $\tau$, we finally get 
\begin{eqnarray}
t=-2M \lim_{x \to 0} \ln (x)-2M.
\end{eqnarray}
The first term goes like $\lim_{x \to 0} \ln (x) \to -\infty$ and, again, this confirms the fact that that the time measured from the outside observer will be very large, i.e., $t\to \infty$, which is in agreement with Eq. (93).  At this point, one can ask whether white holes can be stable remnants? Authors in \cite{Bianchi:2018mml}, argued that such a unitary process  may not violate any
known physics.  This question is outside the scope of the present work, but if there is surrounding matter, most probably the white holes are unstable objects too and collapse again to black holes. According to \cite{Bianchi:2018mml}, there is a difference in the lifetime between the black holes and white holes. The former are described by the law $\tau_{BH} \sim M^3$, and the latter $\tau_{WH} \sim M^4$.  If such a spacetime bounce happens there is a  possibility that strings can increase the size.  This is similar to the so-called fuzzball structure to the black hole, speculated in Ref. \cite{Mathur}.\\

\section{Conclusions}
 In this paper, we  studied the gravitational collapse of matter (dust) under the effect of zero-point length $l_0$. Initially, we have neglected the backreaction effect of the pre-Hawking radiation, then we found that the internal metric of a collapsing star is precisely modeled as a closed FRW universe. Using the modified Friedmann equations and by matching the interior and exterior metrics, we studied the dynamics of the collapsing star and found that the gravitational collapse stops at some minimal scale factor meaning that the particles never hit the singularity.  We argue that such an object emerging at the end of the gravitational collapse are Planck stars hidden inside the event horizon of the black hole with radius proportional to $R_s \sim (GMl_0^{2}/c^2)^{1/3}$. To enhance this conclusion, we found the same result for radius for the Planck star using a free falling observer point of view. 
 
 In the final part of this work we have speculated about the final stage and pointed out two possibilities: (i) First possibility is that black holes (and the Planck stars inside the core of black hole), due to the backreaction effect of the Hawking evaporation decrease their mass to a specific value where there exists an extremal configuration, at this point, the Hawking temperature vanishes, and the resulting object is a Planck-size remnant (a particle). (ii) Second possibility is that the inner core (Planck star), might be unstable. In particular, due to the quantum tunnelling effect, the spacetime can undergo a black hole-to-white hole transition (a bouncing Planck star). We also showed that, from  the outside point of view, the collapse/bounce are viewed in  a very slow motion due to the strong redshift effect. For a stellar mass black holes we have estimated the Planck star radius $10^{-22}$ m, hence it is naturally to expect a generation of gravity weaves along with electromagnetic radiation during the bouncing effect. 
 In the near future, we are planing to study more about the phenomenological aspects of the Planck stars.


\begin{thebibliography}{}
\bibitem{os}
J.~R.~Oppenheimer and H.~Snyder,
  Phys.\ Rev.\  {\bf 56}, 455 (1939).
  
\bibitem{penrose} R. Penrose
Phys. Rev. Lett. 14, 57, 1965

\bibitem{hawkingbh}
S.~W.~Hawking
Nature {\bf 248}, 30-31, (1974); 
S.~W.~Hawking,
Commun. \ Math. \ Phys. \ 43, 199, (1975).

\bibitem{Bardeen} J. M. Bardeen, 
in Proceedings of International Conference GR5, 1968, Tbilisi, USSR, p. 174.

\bibitem{Ayon-Beato:1998hmi}
E.~Ayon-Beato and A.~Garcia,
Phys. Rev. Lett. \textbf{80} (1998), 5056-5059

\bibitem{Haward} S. A. Hayward, 
Phys. Rev. Lett. 96 (2006) 031103

\bibitem{Frolov}
V.~P.~Frolov,
Phys. Rev. D \textbf{94} (2016) no.10, 104056V

\bibitem{Simpson:2018tsi}
A.~Simpson and M.~Visser,
JCAP \textbf{02} (2019), 042;
E.~Franzin, S.~Liberati, J.~Mazza, A.~Simpson and M.~Visser,
JCAP \textbf{07} (2021), 036

\bibitem{Jusufi:2022cfw}
K.~Jusufi,
Annals Phys. \textbf{448} (2023), 169191

\bibitem{Sebastiani:2022wbz}
L.~Sebastiani and S.~Zerbini,
[arXiv:2206.03814 [gr-qc]].

\bibitem{Vagnozzi:2022moj}
S.~Vagnozzi, R.~Roy, Y.~D.~Tsai, L.~Visinelli, M.~Afrin, A.~Allahyari, P.~Bambhaniya, D.~Dey, S.~G.~Ghosh and P.~S.~Joshi, \textit{et al.}
[arXiv:2205.07787 [gr-qc]].


\bibitem{s2}
M.~Cadoni, M.~De Laurentis, I.~De Martino, R.~Della Monica, M.~Oi and A.~P.~Sanna,
[arXiv:2211.11585 [gr-qc]].

\bibitem{r1}
A.~Bonanno, A.~P.~Khosravi and F.~Saueressig,
[arXiv:2209.10612 [gr-qc]].

\bibitem{r2}
A.~Bonanno and F.~Saueressig,
[arXiv:2211.09192 [gr-qc]].

\bibitem{r3}
R.~Carballo-Rubio, F.~Di Filippo, S.~Liberati, C.~Pacilio and M.~Visser,
[arXiv:2212.07458 [gr-qc]].

\bibitem {padma} T. Padmanabhan, Phys. Rev. Lett. 78, 1854 (1997), arXiv:hep- th/9608182.


\bibitem{Nicolini:2019irw}
P.~Nicolini, E.~Spallucci and M.~F.~Wondrak,
Phys. Lett. B \textbf{797} (2019), 134888

\bibitem{Smailagic}
A.~Smailagic, E.~Spallucci and T.~Padmanabhan,
[arXiv:hep-th/0308122 [hep-th]].

\bibitem{Nicolini:2022rlz}
P.~Nicolini,
Gen. Rel. Grav. \textbf{54} (2022) no.9, 106

\bibitem{Gaete:2022une}
P.~Gaete and P.~Nicolini,
Phys. Lett. B \textbf{829} (2022), 137100



\bibitem{k1}
P.~Gaete, K.~Jusufi and P.~Nicolini, Phys. Lett. B \textbf{835} (2022), 137546

\bibitem{kr}
K.~Jusufi,
[arXiv:2209.04433 [gr-qc]].

\bibitem{kgbt}
K.~Jusufi,
doi:10.1088/1674-1137/acaaf4
[arXiv:2206.01189 [gr-qc]].

\bibitem{k2}
K.~Jusufi and A.~Sheykhi,
Phys. Lett. B \textbf{836} (2023), 137621


\bibitem{kt}
K. Jusufi, Phys. Dark Univ. \textbf{39} (2023), 101156.


\bibitem{Pourhassan:2019luf}
B.~Pourhassan, S.~S.~Wani and M.~Faizal,
Nucl. Phys. B \textbf{960} (2020), 115190

\bibitem{Rovelli:2014cta}
C.~Rovelli and F.~Vidotto,
Int. J. Mod. Phys. D \textbf{23} (2014) no.12, 1442026


\bibitem{pls1}
A.~Barrau and C.~Rovelli,
Phys. Lett. B \textbf{739} (2014), 405-409

\bibitem{pls2}
C.~Rovelli and F.~Vidotto,
Universe \textbf{4} (2018) no.11, 127


\bibitem{Bianchi:2018mml}
E.~Bianchi, M.~Christodoulou, F.~D'Ambrosio, H.~M.~Haggard and C.~Rovelli,
Class. Quant. Grav. \textbf{35} (2018) no.22, 225003

\bibitem{pls3}
A.~Barrau, B.~Bolliet, M.~Schutten and F.~Vidotto,
Phys. Lett. B \textbf{772} (2017), 58-62


\bibitem{pls4}
F.~Vidotto, A.~Barrau, B.~Bolliet, M.~Shutten and C.~Weimer,
Springer Proc. Phys. \textbf{208} (2018), 157-163

\bibitem{carlos} Carlos Barcelo, Ral Carballo-Rubio, Luis J. Garay, Class.
Quantum Grav. 34 105007 (2017)

\bibitem{Barcelo:2015uff}
C.~Barcel\'o, R.~Carballo-Rubio and L.~J.~Garay,
JHEP \textbf{01} (2016), 157


\bibitem{Bambi:2013caa}
C.~Bambi, D.~Malafarina and L.~Modesto,
Phys. Rev. D \textbf{88} (2013), 044009

\bibitem{Liu:2014kra}
Y.~Liu, D.~Malafarina, L.~Modesto and C.~Bambi,
Phys. Rev. D \textbf{90} (2014) no.4, 044040

\bibitem{Hajicek:2001yd}
P.~Hajicek and C.~Kiefer,
Int. J. Mod. Phys. D \textbf{10} (2001), 775-780

\bibitem{Kiefer:2019csi}
C.~Kiefer and T.~Schmitz,
Phys. Rev. D \textbf{99} (2019) no.12, 126010

\bibitem{misner}W. C. Hernandez and C. W. Misner, Astrophys. J. 143,
452 (1966).
\bibitem{Mersini-Houghton:2014zka}
L.~Mersini-Houghton,
Phys. Lett. B \textbf{738} (2014), 61-67

\bibitem{lec} 
Ingemar Bengtsson
\newblock{Spherical Symmetry and Black Holes}
\newblock{http://www.fysik.su.se/~ingemar/sfar.pdf}

\bibitem{Chen:2017pkl}
P.~Chen, W.~G.~Unruh, C.~H.~Wu and D.~H.~Yeom,
Phys. Rev. D \textbf{97} (2018) no.6, 064045

\bibitem{Dey}
R.~Dey, S.~Liberati, Z.~Mirzaiyan and D.~Pranzetti,
Phys. Lett. B \textbf{797} (2019), 134828

\bibitem{Mathur} S. D. Mathur, Fortsch. Phys. 53, 793-827 (2005)

\bibitem{LQG} A. Ashtekar and D. Sloan, , Gen. Rel.
Grav. 43, 3619 (2011).


\end{thebibliography}
\end{document}